 \documentclass[final,5p,times,twocolumn]{elsarticle}

\usepackage{amssymb}
\usepackage{amsfonts}
\usepackage{amsmath}
\usepackage{amssymb}
\usepackage{xcolor}

\journal{Physics Letters B}

\begin{document}

\begin{frontmatter}



\title{Chiral transition in a Non-Abelian Quasi-Particle Model with three quark flavours}

\author{E.~P. Politis}
\address{Department of Physics, University of Athens, GR-15874 Athens, Greece}
\author{A. Tsapalis}
\address{Hellenic Naval Academy, Hatzikyriakou Avenue, Pireaus 185 39, Greece}
\address{Department of Physics, University of Athens, GR-15874 Athens, Greece}
\author{F.~K. Diakonos}
\address{Department of Physics, University of Athens, GR-15874 Athens, Greece}

\begin{abstract}

\noindent
We combine the recently introduced Non-Abelian Quasi-Particle Model (NAQPM) for gluons with an ideal Fermi gas of three quark species with the aim to describe the equation of state (energy density vs. temperature) of $2+1$ - flavour Lattice-QCD at zero chemical potential. Allowing temperature dependent masses for the fermions, we show that above a critical temperature $T_c$ the quark mass has to drop rapidly in order to obtain energy density values compatible with the Lattice-QCD results. 
Within this framework, thus, the restoration of chiral symmetry in the system is observed. Furthermore, we demonstrate that the gluon variance --which is a fundamental quantity of the NAQPM-- is strongly correlated to the fermion mass and decreases by orders of magnitude through the transition.  The high temperature phenomenological characteristics of the gluon appear consistent to properties of the perturbative QCD gluon. The model indicates that color deconfinement and chiral symmetry restoration are interrelated and classical configurations of the QCD dynamics play an important role to the criticality of the system.
   
\end{abstract}


\end{frontmatter}


\section{Introduction}
\label{sec1}

\noindent Quasi-particle models often constitute a powerful tool for the simplified description of complex quantum systems. Quasi-particles, firstly introduced by Landau \cite{Landau1957}, are able to encode in their phenomenological characteristics (mass, width {\it{etc}}) complicated and usually non-perturbative many-body interactions leading to collective behaviour of the composite system. Recently, in the example of $Z_2$ Lattice scalar field theory, it was demonstrated that the quasi-particle concept arises from first principles as an effective description away from the critical point \cite{Schweitzer2020}. In the context of strong interacting matter quasi-particle models \cite{QuasiParticles1} are useful since they allow the thermodynamical description even in the regime where Lattice QCD calculations contain large uncertainties. Additionally, when based on suitable physical assumptions, they can be employed as an interpretation tool of the data obtained by tedious Lattice QCD calculations \cite{QuasiParticles1,QuasiParticles2}.

The work presented here is embedded in the framework of quasi-particle approach. It relies on a preceding work \cite{Politis2016} where a non-abelian quasi-particle description of the pure $SU(3)$ Yang-Mills (YM) theory at finite temperature was introduced. Non-linear plane wave solutions of the $SU(3)$ YM equations of motion with arbitrary mass parameter reflecting the scale invariant character of the theory were identified as free quasi-particles. 
The model included transverse and longitudinal gluon modes with quasi-Gaussian distributed masses and contained two temperature dependent parameters: the most probable value of the distribution (mode in statistics terminology) and its variance (actually a width parameter of the distribution closely related to the variance). The temperature dependence
for both transverse and longitudinal gluon modes was obtained from the Lattice calculations of Ref.~\cite{Silva2014} of gluon propagator poles.
The common variance for the transverse/longitudinal modes was utilized as the single free parameter which is able to perfectly reproduce the $SU(3)$-Lattice equation of state \cite{Boyd1996,Borsanyi2012}. Thus, a consistent embedding of the temperature dependence of the gluon mass distribution was possible within this treatment. The NAQPM was based on a statistical mechanical treatment of the gluon system employing microstates which where obtained from periodic solutions of the corresponding non-linear classical equations of motion~\footnote{ The non-linear plane waves used in \cite{Politis2016} were constructed from embeddings of known $SU(2)$ solutions in $SU(3)$. Genuine $SU(3)$ non-linear plane waves have been found in \cite{Tsapalis2018}.}.  
In this way, the quasi-particles include effects of the $SU(3)$ interaction via the mass dependence of the non-linear plane wave amplitude - a feature which is not present in the traditional perturbative treatment of the gluonic states. Allowing distributed mass parameters for the modes and including a variance, on the other hand, addresses the quantum nature of the YM dynamics where quantum fluctuations break dynamically the scale invariance and establish a scale for the propagating modes. {\it{ Thus, within the NAQPM the quantum properties of the YM theory are effectively described via the mass distributions, their most probable values and variances, as well as the use of anharmonic exact plane-wave solutions of the theory as quasi-particle microstates.}}  

Encouraged by the successful description of the thermodynamical properties of the $SU(3)$-system on the Lattice by the NAQPM, we will try in the present work to enrich the model by adding flavours of fermionic components (quarks). Our goal is to describe the Equation of State of $2+1$- flavour Lattice-QCD data at zero baryo-chemical potential with the {\it{Extended}} Non-Abelian Quasi-Particle Model (ENAQPM). The QCD Equation of State (EoS) at zero chemical potential has been computed non-perturbatively via Lattice QCD calculations up to temperatures of $\sim 2$ GeV \cite{Karsch2014, Borsanyi2014, Bazavov2018} for 2+1 quark flavours near their physical masses and extrapolations to the continuum limit. Results are also available for the $N_f = 2+1+1$ theory \cite{Borsanyi2016} and recently calculations have appeared for the high temperature, $3 \le T \le 165$~GeV, regime \cite{Bresciani2025a,Bresciani2025b} allowing contact to perturbative calculations.

We will show in the following that a description of the QCD equation of state for temperatures in the 150-400~MeV regime where the analytic crossover is located, is possible via the inclusion of flavours of non-interacting fermion species with a temperature-dependend mass. 
Indeed, we will demostrate that by fitting the Lattice QCD data for the energy density in \cite{Karsch2014} , a rather intricate correlation between the fermion mass and the gluon mass distribution variance arises as the temperature increases and the system experiences a fast increase of the energy density. In fact, the ENAQPM captures the restoration of the chiral symmetry in the quark sector and in addition, it indicates the emergence of (almost) conventional particle properties (zero mass, reduced variance)  for the gluon in the high temperature regime.  

The remaining article is organized as follows: in section II we describe the ENAQPM and the calculation of the relevant thermodynamic quantities.
In section III we describe the algorithmic fitting process and we investigate the temperature dependence of the quark and gluon microstate properties that allow a succesful description of the Lattice-QCD data. We show how we are led to a picture where the mass of the fermionic flavors rapidly decreases as the temperature increases while at the same time the gluon microstates properties are altered drastically with the rapid decrease of the variance of the mass distribution. In this high temperature regime the pure $SU(3)$-based gluon description fails to reproduce the Lattice-QCD data and the gluon parameters have to be significantly modified relative to the pure $SU(3)$ scenario in order  to achieve a satisfactory description of the Lattice-QCD data. We also investigate how our results vary with the inclusion of two or three quark flavors in the thermodynamics of the model. 
Finally, in section IV we present our concluding remarks and aspects for future investigations.

\section{The Extended Non-Abelian Quasi-Particle Model} 
\label{sec2}

\noindent
The extended non-abelian quasi-particle model considers a composite system of two ideal gases: the bosonic gluon gas, described by the NAQPM  \cite{Politis2016} and the fermionic quark gas. Self-interaction effects of the gluonic sector are incorporated into the temperature dependent parameters of the NAQPM so that the pure $SU(3)$ -Lattice thermodynamics are perfectly reproduced. We refer in the following to this gluonic component as the ``pure non-abelian gluon''. Initially, we assume that the interaction between gluons and quarks is captured by the temperature dependence of the quark mass. We will see in the subsequent analysis that this picture works only for temperatures in a restricted region up to a temperature around $230$ MeV. For even higher temperatures the quark mass saturates at a very small value and the only parameters which can absorb the interactions in the quark-gluon system are the temperature dependent most probable gluon mass and its variance. As we argue below, in this high temperature region the gluon ceases to be the pure non-abelian gluon. 

We start the presentation of the ENAQPM recalling, for completeness, the formal expression for the energy density component of the gluons, as dictated by the NAQPM. More details concerning the basic features of the NAQPM can be found in \cite{Politis2016}. Denoting with $tr, lo$ subscripts the transverse and longitudinal gluon modes respectively, the energy density of the model is given by ($k_B = 1$):
\begin{equation}
 \epsilon_g= \epsilon_{g,tr} + \epsilon_{g,lo}
 \label{eq:eq1a} 
\end{equation}
with corresponding contributions:
\begin{multline}
\epsilon_{g,tr}= \frac{64 \pi \cdot N_{tr} \cdot T^4}{{\mathcal{P}}^3} \sum _{l=1} ^ {\infty} \int_0^{\infty} dm \quad \exp \left[-\frac{(m-\mu_{tr})^2}{2 {\sigma}^2}\right] \cdot \frac{1}{l^4} \\
\cdot \left[ 3\left(\frac{m}{T} \, l \right)^2 \, K_2 (\frac{m}{T} \, l ) + \left(\frac{m}{T} \, l \right)^3 \, K_1 (\frac{m}{T} \, l ) \right] \nonumber 
\end{multline}

\begin{multline}
\epsilon_{g,lo}= \frac{32 \pi \cdot N_{lo} \cdot T^4}{{\mathcal{P}}^3} \sum _{l=1} ^ {\infty} \int_0^{\infty} dm \quad \exp \left[-\frac{(m-\mu_{lo})^2}{2 {\sigma}^2} \right] \cdot \frac{1}{l^4} \\
\cdot \left(\frac{m}{T} \, l \right)^3 \cdot K_1(\frac{m}{T} \, l)
\label{eq:eq1}
\end{multline}

\noindent with $K_\nu$ the modified Bessel functions \cite{Abramowitz1987}, $\mu_{tr}$ and $\mu_{lo}$ the most probable gluon transverse and longitudinal masses respectively and $\sigma$ the gluon mass variance. The parameters $\mu_{tr}$, $\mu_{lo}$ and $\sigma$ all depend on the temperature $T$. We use a single variance, for both longitudinal and transverse masses, to minimize the number of free parameters. This is a reasonable assumption since the variance itself is associated within our model with the presence of quantum fluctuations and both types of modes are indeed expected to fluctuate in a correlated manner. 
It has been checked in \cite{Politis2016} that allowing independent variances for the masses of the gluon modes does not influence the results. 
In addition, the factor $\mathcal{P}=5.244$ is the period of the anharmonic plane-wave solutions (\cite{Politis2016}) and $N_i$ the normalization factors for the quasi-Gaussian mass distributions:
\begin{multline}
N_i \equiv N_i(\mu_{i},\sigma)=\frac{1}{\sigma} \sqrt{\frac{2}{\pi}}\cdot \left({1+Erf(\frac{\mu_i}{\sqrt{2}\,\sigma})}\right)^{-1}~~~~~;~~~~~i=tr,~lo 
\nonumber \\
\end{multline}
Employing Eq.~(\ref{eq:eq1a}) one can perfectly describe the pure $SU(3)$-Lattice data \cite{Boyd1996,Borsanyi2012} while fixing at the same time $\mu_{lo}(T)$ and $\mu_{tr}(T)$ to the values obtained in the Lattice calculation of Ref.~\cite{Silva2014}, provided that $\sigma(T)$ takes the values obtained by the fit in \cite{Politis2016}. With these choices the gluon component in Eq.~(\ref{eq:eq1}) is identified as the ``pure non-abelian gluon''. Notice that in this case $\sigma$ attains a value of $\approx 3$ GeV at the critical temperature and decreases with increasing temperature, reaching a minimum at $t=\frac{T-T_c}{T_c} \approx 0.3$ and increasing almost linearly with $T$ beyond this minimum \cite{Politis2016}.

In a completely analogous way (\cite{Kapusta}, \cite{Shuryak1977}), the energy density of a fermionic (spin-1/2, matter-antimatter and $SU(3)$-colored) component of mass $m_f$ is derived:
\begin{align}
\epsilon_{f}= 4\cdot 3 \cdot \int  \frac{d^3 \vec{k}}{(2\pi)^3} \, \frac{\omega(\vec{k}, m_f)}{e^{\beta \omega} + 1}  \;\;\;, \;\; \omega(\vec{k}, m_f) = \sqrt{\vec{k}^2+m_f^2}
\end{align}
Substituting $k = |\vec{k}|=m_f \sinh(t)$ and using known properties of the Bessel functions, the integration is performed and leads to:
\begin{equation}
\epsilon_{f}=\frac{6 \, T^4}{\pi ^2} \sum_{l=1} ^{\infty} (-1)^{l+1}\cdot \frac{1}{l^4} \cdot \left[ 3\left(\frac{m_f}{T}  l \right)^2  K_2 (\frac{m_f}{T}  l ) + \left(\frac{m_f}{T}  l \right)^3  K_1 (\frac{m_f}{T}  l ) \right] 
\label{eq:eq2}
\end{equation}
with $m_f = m_f(T)$ the temperature dependent quasi-particle quark mass. Our aim is to describe the Lattice-QCD thermodynamics at zero chemical potential for the case of two light and one heavier quark flavours. To simplify our quasi-particle description, we reduce the parameter space using a single quasi-particle quark mass parameter consistent with the assumption of three mass-degenerate quark flavours. In fact, approaching the transition regime from the hadron phase (low temperatures), one expects the quark masses to be close to the constituent values which can be taken approximately as degenerate since: $m_u \approx m_d \approx 340$ MeV and $m_s \approx 490$ MeV \cite{Griffiths2008} with a ratio of $\sim 1.4$. Within our treatment we arrive at a (degenerate) quark quasi-particle mass at the critical temperature $m_q(T_c)\sim 490$ MeV. 
We assume a relativistic hadron gas description \cite{Cleymans1993} to be valid within the hadronic phase, in consistency with Lattice-QCD \cite{Bazavov2012} calculations. In the relativistic hadron gas description the hadron masses, as well as the associated constituent quark masses do not depend on the temperature. Our study on the other hand, is focused on the deconfined regime for which we adopt the quasi-particle picture. In this regime the total energy density of the ENAQPM is given by:
\begin{equation}
\epsilon_{tot}=\epsilon_{g,tr}+\epsilon_{g,lo}+ 3\cdot \epsilon_{f}
\label{eq:eq3}
\end{equation}
since its components are assumed as non-interacting. From the basic thermodynamic relations for energy density, temperature, pressure and entropy density $s$
\begin{equation}
\epsilon =  T s - P  \,\,\, ; \,\,\, s = \frac{\partial P}{\partial T} \Rightarrow
\end{equation}
\begin{equation}
\epsilon =  T  \frac{\partial P}{\partial T} - P \Rightarrow \epsilon = T^2 \frac{\partial}{\partial T} \left(\frac{P}{T}\right)
\label{eq:eq3a}
\end{equation}
we obtain the pressure via integrating Eq.~(\ref{eq:eq3a}) with the energy density for the ENAQPM defined in Eq.~(\ref{eq:eq3}):
\begin{equation}
\frac{P}{T}=\frac{P_o}{T_o} +\int_{T_o} ^{T} dT ~\frac{\epsilon_{tot}}{T^2} \,.
\label{eq:eq4}
\end{equation}
$T_o$ and $P_o$ denote initial (reference) values for the temperature and pressure respectively such that $P(T_o)=P_o$. The thermodynamic consistency of the above relations in the presence of a temperature-dependent quasiparticle mass has been discussed in \cite{Bannur2012}.
The trace anomaly, $\Delta = \epsilon - 3 P$, which is the trace of the energy-momentum tensor of the theory, is also thermodynamically related to pressure via
 \begin{equation}
\frac{\Delta}{T^4}=  T \frac{\partial}{\partial T} \left(\frac{P}{T^4}\right)  \,.
\label{eq:eq4a}
\end{equation}
With the above relations, all thermodynamical quantities of interest are computed via the energy density of the model, which in turn depends on the fermion mass in the quark sector as well as the gluon variance and the most probable transverse and longitudinal masses in the gluon sector. The interaction of the modes -both the self-interaction of the non-abelian fields as well as the quark-gluon dynamics is assumed to be included principally in the temperature dependence of these phenomenological parameters of the ENAQPM.

\section{Energy density of ENAQPM vs. (2+1) Lattice-QCD results}

\subsection{Constraining the Model}

\begin{figure}[htb]
\centerline{\includegraphics[width=\columnwidth]{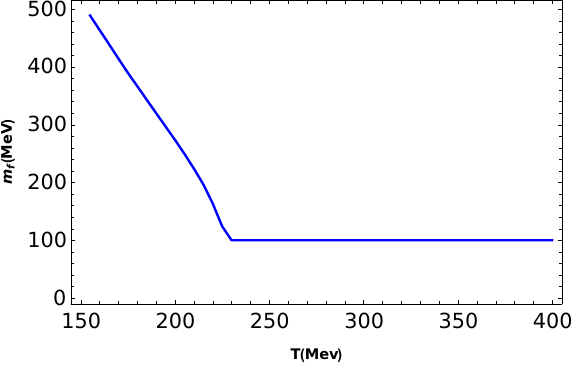}}
\vspace{3mm}
\centerline{\includegraphics[width=\columnwidth]{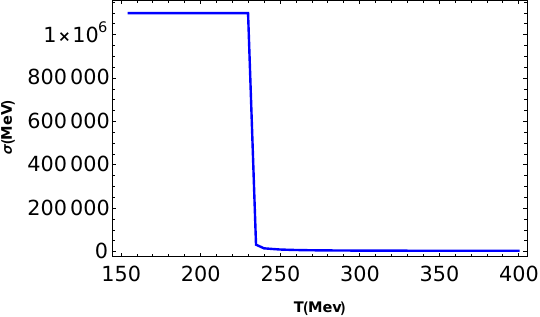}}
\caption{(a) The temperature dependence of the quark mass, $m_f(T)$, is shown for the three-flavor ENAQPM. In the $[155, 230]$ MeV interval the mass is the fitting parameter with the gluon variance fixed to a constant.
   In the $[230, 400]$ MeV interval, the quark mass is kept fixed to the 100 MeV value. The fitting procedure is described in Subsection 3.1.
  (b) The temperature-dependence of the gluon variance, $\sigma(T)$,  is shown for the three-flavor ENAQPM. In the $[155, 230]$ MeV interval the variance is kept fixed, while in the $[230, 400]$ MeV interval is the fitting parameter.  }
\label{fig:fig1}
\end{figure}

\noindent
In this section we present the determination of the essential parameters of the ENAQPM from the fitting of the energy density of the model (Eq.~\ref{eq:eq3})) to the energy density data for (2+1) Lattice-QCD at zero chemical potential given in \cite{Karsch2014}. Since the estimated transition temperature from the specific heat analysis of the data  is around $155$~MeV, we will restrict our model to the data available in the $[155,400]$~MeV region assuming the validity of the Hadron Resonace Gas in the hadronic region.    
Our goal is to achieve a uniform description of the energy density before and after the peak detected in the trace anomaly data around 200~MeV and provide a robust determination of the behavior of the essential parameters of the model. To that end we start our fitting procedure at the highest temperature data point available which is at 400~MeV. We firstly check the hypothesis if it is possible to fix the gluonic contribution to the pure non-abelian gluon characteristics determined from the non-abelian QPM in \cite{Politis2016} having at the same time a small quark mass for the three flavors in the 10-100~MeV range
(compatible to the onset of chiral symmetry restoration in this temperature regime).    
The fitting algorithm (as described also in \cite{Politis2016}) performs a Monte-Carlo search such that the lattice data is fitted within $10^{-4}$ accuracy or higher~\footnote{In fact the results are quite robust. Changing the convergence criterion from $10^{-4}$ to $10^{-3}$ leads to a change in the second decimal digit for the parameter $\sigma$.}. We observe that it is impossible to fit the energy density derived from the (2+1) Lattice-QCD calculations in \cite{Karsch2014} with the pure non-abelian gluonic parameters and a temperature dependent quark mass for all three flavors. 
The influence of the quark mass at $T = 400$~MeV is saturated and the fermionic contribution to the energy density is minimally affected --at the $10^{-3}$ level -- by the variation of the quark mass in the 10-100~MeV~\footnote{More precisely, altering the fermion mass from 100 to 10~MeV affects the energy density at the $10^{-2}$ level for $T=155$~MeV and  $10^{-3}$ level for $T=400$~MeV. Altering from 10~MeV to 1~MeV affects the the energy density between $10^{-3}$ and $10^{-5}$ respectively.}.
It becomes therefore clear that the gluonic contribution to (\ref{eq:eq3}) cannot be described by the variance and most probable longitudinal and transverse gluon masses as estabished in \cite{Politis2016}. We furthermore notice that the influence of the most probable gluon masses is not important in this regime and the leading behavior is determined by the gluon variance $\sigma(T)$. We, thus, proceed with the following scheme: we set the most probable longitudinal and transverse gluon masses, $\mu_{lo}, \mu_{tr}$ in (\ref{eq:eq1}) to zero maintaining the gluon variance $\sigma(T)$ as the dominant and only fitting parameter in the gluon sector. In this context, the parameter $\sigma(T)$ becomes proportional to the precise variance --in the probability sense-- of the gluon mass distribution. 
This is because $\sigma(T)$ is now the only scale dependent parameter of the distribution and the mean mass and variance of the distribution are necessarily proportional to $\sigma(T)$ with proportionality constants that can be computed. Thus, we will continue referring to $\sigma(T)$ as the `variance'  of the model since it is essentially the only important parameter for all statistical averages of the gluon mass distribution.    
In the fermion sector we fix the quark masses to the 100~MeV value for all three flavors. This approach determines an initial fitting value for $\sigma(T=400~\text{MeV})$ 
 and the algorithm proceeds to sequentially fitting the energy density by lowering the temperature in steps of 5~MeV. As described in \cite{Politis2016}, having located the optimal value $\sigma(T_i)$ for the temperature $T_i$, we estimate the value $\sigma(T_{i+1})$ by exploring a region centered at $\sigma(T_i)$ and extending        
up to $50\%$ around it in each direction for an (ordered) sequence of temperatures $T_i, i=1,2,\dots,N$ with $T_{i+1} < T_i$.
This procedure converges fast to the optimal value, which is important due to the extensive CPU time required for the performance of the numerical integrations in  (\ref{eq:eq1}).  We observed that the fitting proceeded smoothly all the way down to the temperature value of 230~MeV where the fitted value of $\sigma$ presents a fast increase and the fit becomes unstable. This is clearly shown in Fig.~\ref{fig:fig1}(b) for the branch of data in the  
$[230, 400]$ MeV range of temperatures. We conclude that around 230~MeV the fermion mass starts to becomes important for the model and thus we now 
 fix the gluon variance to the value of $3\times 10^6$ as determined by the last fit at 230~MeV and allow the fermion mass as the free fitting parameter of the algorithm. The fit proceeds with the 
 sequential fitting of the energy density in decreasing steps of 5~MeV for the temperature all the way down to the last available data point at $T=155$~MeV with the fermion mass ~$m_f(T)$ constantly increasing. At the last fitting point at $T=155$~MeV the fermion mass has reached a value close to 490~MeV.
 This approach produces the plot  in Fig.~\ref{fig:fig1}(a) for the branch of data in the $[155, 230]$~MeV range of temperatures. From the two stages of fitting as described above and shown collectively in Fig.~\ref{fig:fig1} we conclude that 
there are two distinct regimes of thermodynamical behavior for the ENAQPM. First, the  $[155, 230]$~MeV regime where a high value for the gluon variance and a smoothly (almost linearly) decreasing value for the fermion mass can describe perfectly the Lattice QCD data. Second, the $[230, 400]$~MeV range where a small fermion mass and a small, smoothly descreasing gluon variance describe the data. In both regions, the quality of fit is high (at the level of $10^{-8}$) as shown from the energy density plot in 
Fig.~\ref{fig:fig3}(a).

 \subsection{Exploring the fermion mass - gluon variance relation} 

\begin{figure}[htb]
\centerline{\includegraphics[width=\columnwidth]{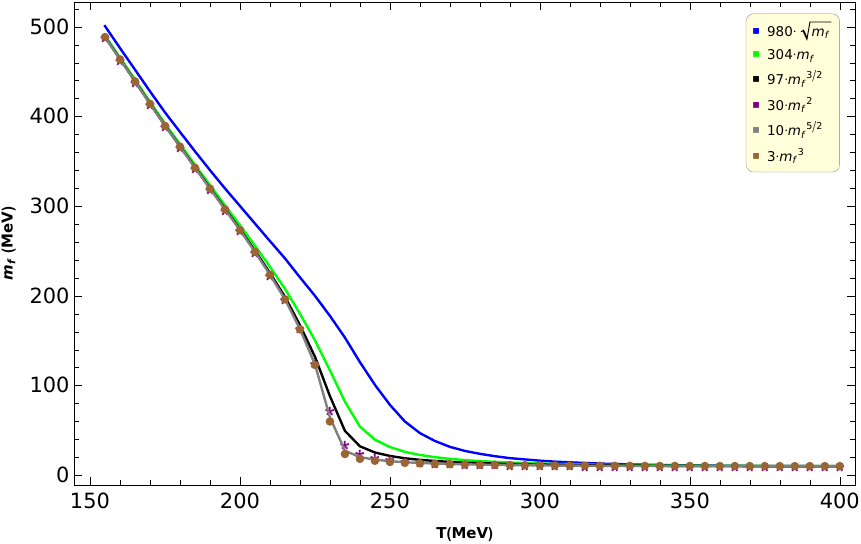}}
\vspace{3mm}
\centerline{\includegraphics[width=\columnwidth]{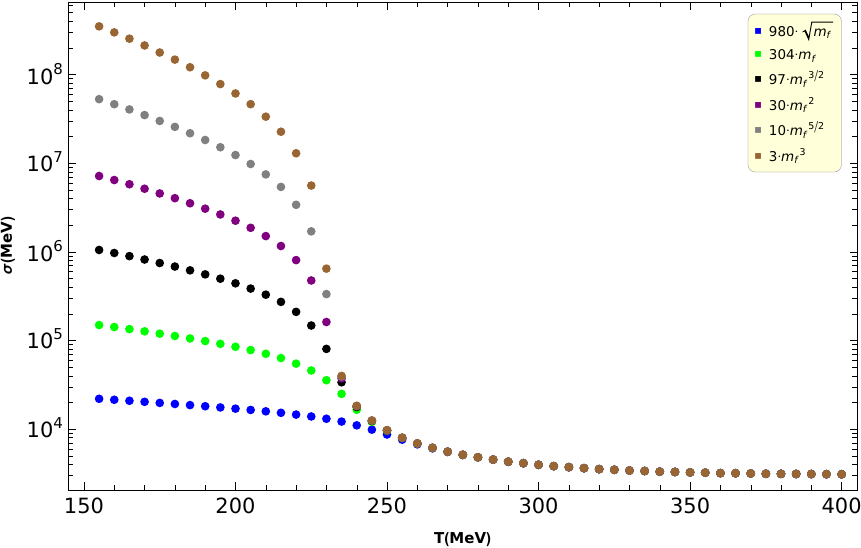}}
\caption{(a) The temperature dependence of the quark mass, $m_f(T)$, is shown for the family of monoparametric fits in the $[155, 400]$~MeV range for the three-flavor ENAQPM. The gluon variance, $\sigma(T)$,  is analytically related to the quark mass by the formulas in the upper right part of the plot.  
(b) The temperature-dependence of the gluon variance, $\sigma(T)$, as determined by the analytic formulas from the fitted quark mass. 
}
\label{fig:fig2}
\end{figure}

\noindent
The main conclusion of the previous subsection is that the ENAQPM can describe satisfactorily the energy density of  (2+1)-flavor Lattice-QCD with one dominant temperature dependent parameter in the $[155, 400]$ MeV interval with the following characteristics:
\begin{itemize}
\item Above the temperature value of 230~MeV the fermion mass has to be kept at small values. Its exact value does not influence the total energy density value and variations of its value from $m_f \approx 10$~MeV to the $m_f \approx 100$~MeV value alter the energy density by less than $1\%$ at the 230~MeV regime.
\item  One has to abandon the pure non-abelian gluon characteristics (\cite{Politis2016}) in order to achieve a satisfactory description of the Lattice-QCD data above the 230~MeV temperature. In this region, the only important parameter that affects the overall behavior of the energy density of the model is the gluon variance. The most probable longitudinal and transverse gluon masses, $\mu_{lo}, \mu_{tr}$ introduced in the quasi-Gaussian distributions in (\ref{eq:eq1}) turn out not to affect importantly the behavior of the energy density. In order to avoid multiparameter fits which are more unstable we have set them to zero. 
\item Below the 230~MeV temperature the system is described satisfactorily via a temperature-dependent and decreasing fermion mass for all three flavors while the gluon variance presents a fast increase as the temperature is lowered through the 230~MeV value.
\end{itemize}
Based on the above observations we investigate in this subsection the possibility of an analytic relation connecting the fermion mass with the gluon variance that can capture simultaneously the behavior of both parameters with a unique description in the whole temperature range $[155, 400]$ MeV where data are available. 
In particular, we examined power-law relations of the type
\begin{equation}
\sigma(T) = C \cdot m_f(T)^{\nu}
\label{eq:eq5}
\end{equation}
for various values of the exponent $\nu$ and an accordingly determined constant $C$. For the exponent values of $\nu = 1/2, 1, 3/2, 2, 5/2$ and $3$,  the energy density can be fitted precisely with the fermion mass $m_f(T)$ as the only fiting parameter of the model in the whole $[155, 400]$ MeV range provided that the gluon variance is constrained by Eq.~(\ref{eq:eq5}). In Fig.~\ref{fig:fig2}(a) the fitted values for the temperature-dependent fermion mass are shown for the set of analytic relations given in the upper right corner. The terminal value for the fermion mass at $T=400$~meV is fixed at 10~MeV and its effect on the energy density is below the $10^{-3}$ level at this point.
The corresponding values for the gluon variance $\sigma(T)$ as determined via Eq.~(\ref{eq:eq5}) are shown in Fig.~\ref{fig:fig2}(b) for the family of selected exponents $\nu$. We checked also that the fitting works equally well for smaller and larger values of the exponent $\nu$ than the ones in the interval $1/2 - 3$ shown above. In fact, even for $\nu = 1/16$, which is the smallest one we tried, the fitting works with similar accuracy, the fermion mass decreases with the same characteristics as in Fig.~\ref{fig:fig2}(a) and the gluon variance presents a milder decrease with respect to the choises in Fig.~\ref{fig:fig2}(b).   

\begin{figure}[htb]
\centerline{\includegraphics[width=\columnwidth]{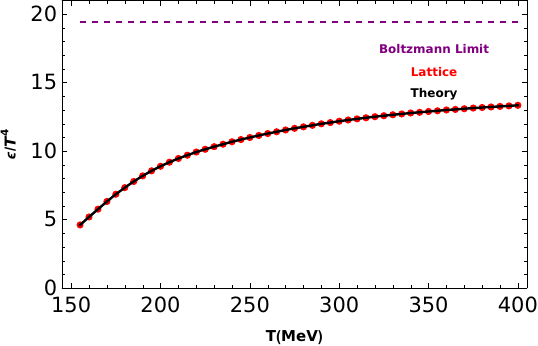}}
\vspace{3mm}
\centerline{\includegraphics[width=\columnwidth]{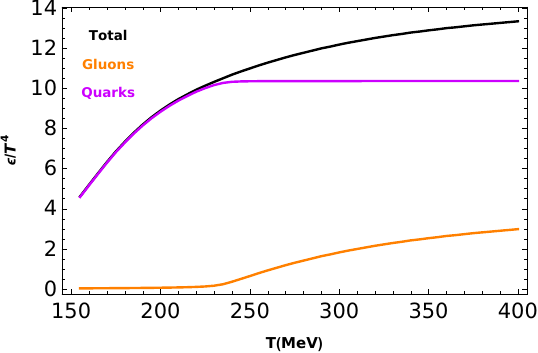}}
\caption{(a) The energy density of 2+1 Lattice-QCD \cite{Karsch2014} as a function of temperature for $155 < T < 400$ MeV (red circles). 
The energy density of the ENAQPM is shown by the solid line. The Boltzmann limit for the massless anharmonic gauge bosons plus fermions is included in the plot by the dotted line. 
(b) The ENAQPM total energy density is shown by the black line along with the gluonic contribution (orange line) and the three quark flavors contribution (magenta line).}   
\label{fig:fig3}
\end{figure}

\noindent
The energy density of the ENAQPM versus temperature is shown in  Fig.~\ref{fig:fig3}(a) along with the Lattice QCD data~\cite{Karsch2014}. The accuracy of the fitted Lattice data is at the level of $10^{-8}$ for all the monoparametric fits shown 
in Fig.~\ref{fig:fig2}. The model data for the relation $\sigma(T) =304\; m_f(T)$ are shown in the plot but no difference would be visible to the eye if we used any other relation. The non-interacting Boltzmann limit line is included in the plot. Each massless gauge boson which is a harmonic wave solution contributes $\pi^2/15$ in this limit but we utilize the non-abelian plane waves of \cite{Politis2016} which are anharmonic with a period of $\mathcal{P}=5.244$ and thus the gauge boson Boltzmann limit is modified to $\pi^2/15 \times \left(\frac{2\pi}{\mathcal{P}}\right)^3 \simeq 1.132$.    
On the other hand, each Dirac fermion contributes, as usual,  $7\pi^2/60$ to that limit. Thus, the Boltzmann limit line for the ENAQPM is at $\sim19.4$, higher than the standard value obtained counting eight massless non-interacting gluons which would be around $15.6$.
The individual contribution of the gauge boson and the quark species to the total energy density of the model is also presented in Fig.~\ref{fig:fig3}(b). We observe that the contribution of the fermion degrees of freedom increases fast in the $[155, 230]$~MeV
interval and around the 230~MeV temperature value  it becomes saturated. Since the fermions are non-interacting and their mass is small for T above 230~MeV, their mass contribution is minimal to the energy density and the fermionic part saturates to the massless free fermion contribution limit which is around $3 \times 3 \times 7\pi^2/60 \sim  10.36$.  On the other hand, the gauge boson contribution is small below 230~MeV and after this value the energy density increase of the full theory is exclusively due to the gluonic contribution. This behavior is related to the gluon variance which controlls the energy density of the model above 230~MeV. The gluonic contribution to the model is attributed to the massive non-abelian plane wave solutions of the $SU(3)$ theory. Owing to the scale invariance of the classical YM theory, all scales participate to the partition function of the model with a quasi-Gaussian invariant mass distribution which peaks at zero and has the variance $\sigma(T)$ as the principal non-perturbative parameter collectively describing the quantum interaction and thermodynamical properties of the model. Within the ENAQPM, we arrive at a picture where in the confining regime the non-abelian plane waves --which are the saddle point solutions of the YM system (\cite{Politis2016})-- propagate with masses that obtain a large range of values (up to hundrends of GeV if we adopt the linear relation  $\sigma(T) =304\; m_f(T)$). These modes are responsible for the gluonic contribution to the energy density which remains small in the confining regime. As we approach the transition temperature around 230~MeV the variance of the gluon modes decreases rapidly. We interpret this behavior as a signal that the average mass of the excited gluon modes decreases rapidly above the transition (two orders of magnitude with the linear relation) and thus we enter the deconfinement regime where the gluons propagate to larger distances with the same energy. The energy density released by these modes is responsible for the increase of the total energy density of the full theory as demonstrated by the Lattice QCD data. This is a picture closer to the standard perturbative view where due to the weakening of the gauge interaction as the temperature increases, gluons can propagate to larger distances and thus a transition (or more precisely crossover) to a phase with the Quark-Gluon Plasma (QGP) characteristics takes place.   

\subsection{Other Thermodynamical quantities} 

\begin{figure}[htb]
\centerline{\includegraphics[width=\columnwidth]{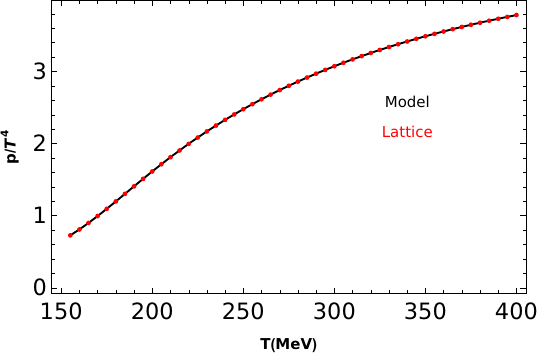}}
\vspace{3mm}
\centerline{\includegraphics[width=\columnwidth]{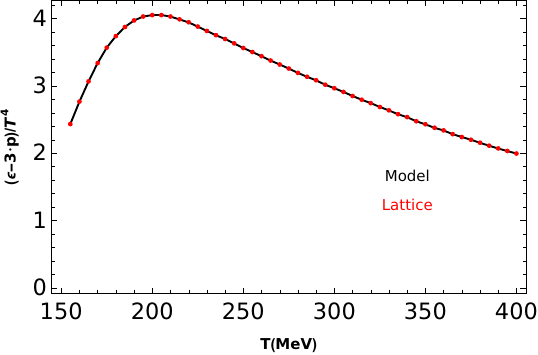}}
\caption{(a) The pressure of 2+1 Lattice-QCD \cite{Karsch2014} as a function of temperature for $155 < T < 400$ MeV (red circles) versus the ENAQPM (solid line).
(b) The trace anomaly, $\Delta= \epsilon-3 P$, for the ENAQPM (solid line) versus the Lattice-QCD data. The peak defining the (pseudo-) transition temperature is clear at the 200~MeV value.  
}
\label{fig:fig4}
\end{figure}

\noindent
Having established the basic contributions to the energy density in the previous subsection, the study of other thermodynamical quantities becomes feasible within the ENAQPM. Initially, the pressure is obtained from the integration in Eq.~(\ref{eq:eq4}). In Fig.~\ref{fig:fig4}(a) we present the calculated values for the pressure over $T^4$ as determined via Eq.~(\ref{eq:eq4}) for our model versus the corresponding Lattice data in \cite{Karsch2014}. Numerical agreement to a high level of accuracy is expected and observed here also.This follows from the accuracy of the energy density data fit and the fact that in both the Lattice and our approach the pressure is computed via integration formulas.
 The trace anomaly,
$\Delta = \epsilon - 3 P$, being the trace of the energy-momentum tensor is of particular interest since it is related on general arguments to the breaking of scale symmetry. From Eq.~(\ref{eq:eq4a}) it is connected to the slope of $p/T^4$  and indeed from Fig.~\ref{fig:fig4}(a) a clear change of slope is evident around a temperature of 200~MeV. In Fig.~\ref{fig:fig4}(b) the Lattice data for the trace anomaly are plotted versus the corresponding values for the ENAQPM. Notice that in our model, $\Delta$ is obtained via a simple subtraction from energy density and pressure data while in the Lattice methodology $\Delta$ is the primarily computed quantity via Euclidean field theory ensembles. 
The peak for this quantity is determined at $\sim204$ ~MeV from this plot in \cite{Karsch2014}. Within our model, the behavior of this plot reflects the change of scales that takes place in the
$[155-230]$~MeV temperature region. Below 155~MeV the confined QCD matter consists of distinct hadron states (ranging from the 140~MeV pion mass level and above) and gluons confined within the size of those hadronic states -- thus again with energy scales at the 200~MeV range and above due to the small size $( < 1\;\text{fm})$ of the hadron states. Within the ENAQPM this picture is captured by three constituent quark flavors with a mass around 500~MeV and a quasi-Gaussian distribution of gluon states with a high value of variance indicating that the average mass-scale of these excitations is also large (hundrends of GeV if we adopt the linear relation $\sigma(T) \sim m_f(T) )$. As the temperature increases, and up to the 230~MeV range, we experience
 a prevalent change of scales with the quarks switching to the current mass values and the gluon variance reducing by orders of magnitude. Within our model, the fermion mass and the gluon variance are the only parameters which incorporate the interaction and thermal effects of QCD. In consequence,  they can also be viewed as the parameters determining the full correlation length behavior of the theory since they connect analytically the hadronic and the QGP regimes. Above 230~MeV,  the fermionic energy density component is saturated and the energy release comes dominantly from the gluonic states which have obtained now a much smaller average mass and can propagate to much larger distances, in agreement with the expectations from high-temperature pertrurbative field theory.     
 This picture is consistent for all the monoparametric fits displayed in Fig.~\ref{fig:fig2}. The energy density, pressure and trace anomaly as shown in Figs.~\ref{fig:fig3}(a),  \ref{fig:fig4}(a) and \ref{fig:fig4}(b) respectively are fitted to an accuracy of $10^{-6}$ or better for all the selected exponents. 

\noindent

\begin{figure}[htb]
\centerline{\includegraphics[width=\columnwidth]{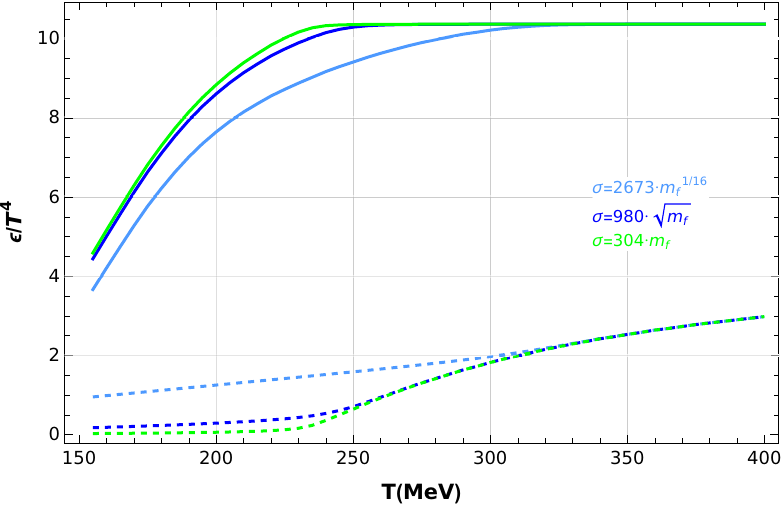}}
\vspace{3mm}
\centerline{\includegraphics[width=\columnwidth]{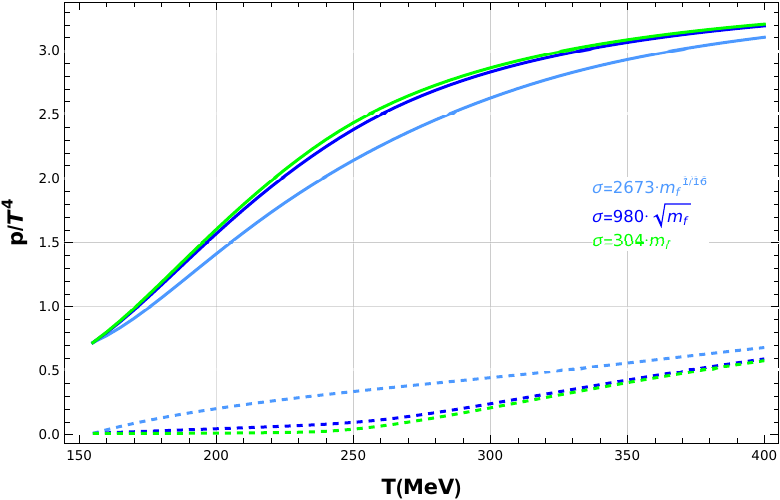}}
\vspace{3mm}
\centerline{\includegraphics[width=\columnwidth]{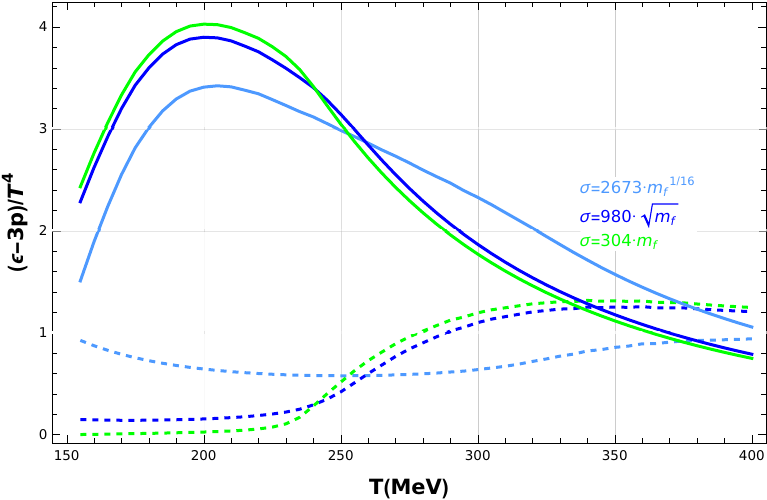}}
\caption{(a) The fermionic contribution (solid lines) and the gluonic contribution (dotted lines) to the energy density of the ENAQPM is displayed for a selection of variance - quark mass relations with specific exponents distinguished by color.
(b) The fermionic and gluonic contributions to pressure with line and color selections as in (a). (c) The trace anomaly for the individual contributions as in (a)-(b). 
}
\label{fig:fig5}
\end{figure}

\noindent
In Fig.~(\ref{fig:fig5}) we examine the separate contributions of the fermionic and gluonic components to these quantities. For a selection of exponents $\nu = 1/16, 1/2 $ and 1 we display with solid lines the quark contribution and with dotted lines the gluon contribution to: (a) the energy density , (b) the presssure, and (c) the trace anomaly (all in units of $T^4$).   For all cases we notice a very similar behavior for these thermodynamic quantities indicating that the rapid change of scales as determined by the fitted quark mass and the associated gluon variance from Eq.~(\ref{eq:eq5}) is a robust behavior of our quasi-particle model. In addition, we observe that the fermionic contribution is the one dominating the behavior of the total thermodynamic 
functions in the crossover region of $155-230$~MeV. 

\begin{figure}[htb]
\centerline{\includegraphics[width=\columnwidth]{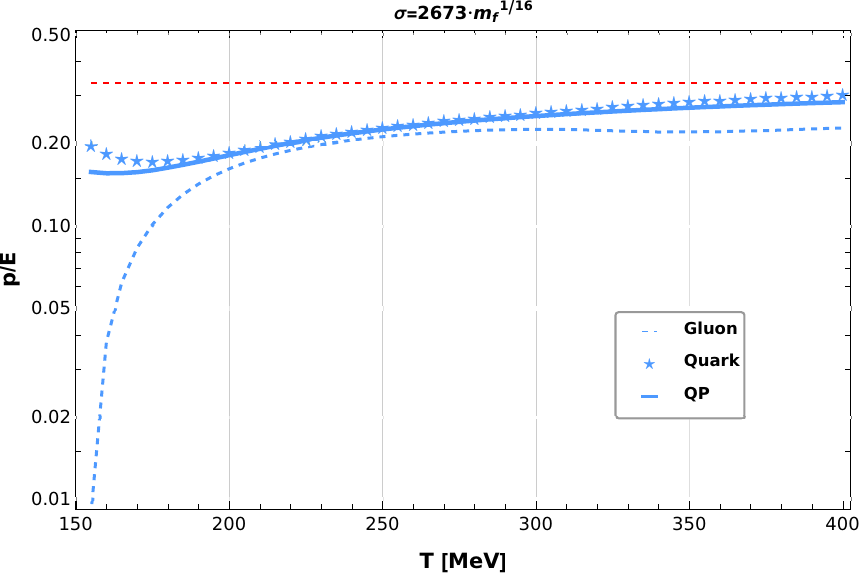}}
\vspace{3mm}
\centerline{\includegraphics[width=\columnwidth]{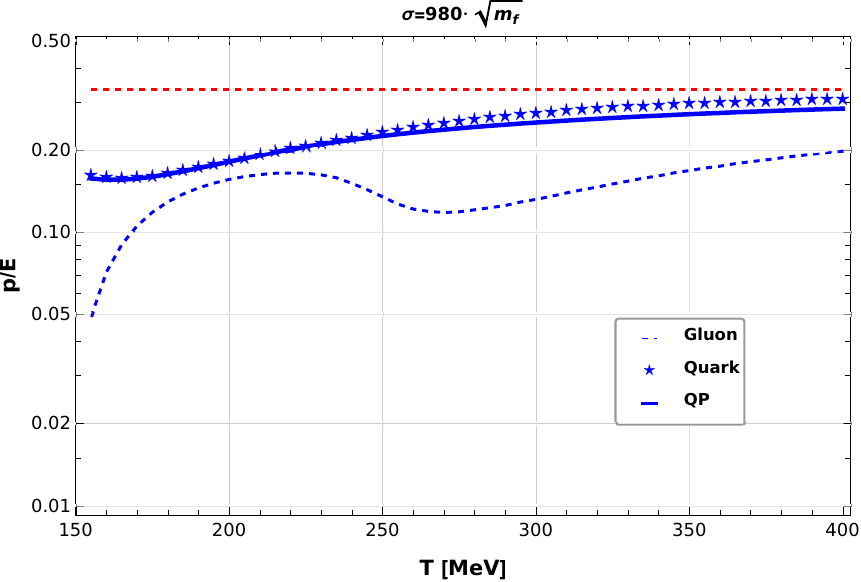}}
\vspace{3mm}
\centerline{\includegraphics[width=\columnwidth]{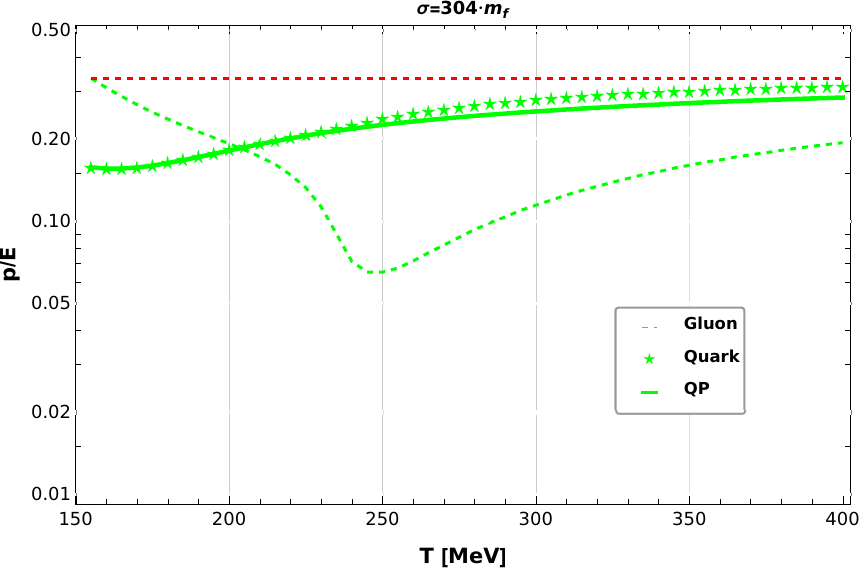}}
\caption{Plots of the pressure over energy ratio for the full ENAQPM (solid line), quark flavors contribution (stars) and the gluon contribution (dotted line) for three different 
anzatses of the $\sigma \sim m^{\nu}_f$ relation. The red horizontal line at the $1/3$ value denotes the massless relativistic limit.   (a) The $\nu=1/16$ exponent. 
(b) The $\nu=1/2$ exponent. (c) The $\nu=1$ exponent. 
}
\label{fig:fig6}
\end{figure}
It is of interest also to examine specifically the pressure over energy ratio behavior for the ENAQPM in full and for the separate modes contribution. 
At the free massless limit it approaches the value of $1/3$ (for both bosonic and fermionic modes), remaining below this value for the massive regime and close to zero at the non-relativistic limit. While the energy contribution for each sector is well defined in the model from Eqs.~(\ref{eq:eq1}), (\ref{eq:eq2}) and (\ref{eq:eq3}), the 
calculation of pressure via the thermodynamically consistent relation (\ref{eq:eq4}) requires an initial condition at a given temperature $T_0$ (taken to be 155~MeV). For the full model the value $P_0$ is provided by the corresponding Lattice calculation. For the individual quark/gluon contributions, 
since we lack an estimation at $T_0$, we assign a percentage $\lambda P_0$ (with  $0 \le \lambda \le 1$ ) to the fermions and $(1-\lambda) P_0$ to the gluons. We varied the values of $\lambda$ for each exponent 
$\nu$ and plotted the associated ratio $p/\epsilon$. We noticed that for small values of $\lambda$ (i.e. gluons dominating the pressure of the plasma) the ratio $p/\epsilon$ can become larger than $1/3$. This behavior is inconsistent with the non-interacting quasi-particle assumption and thus we constrain $\lambda$ by requiring  $p/\epsilon < 1/3$ for the values at $T=155$~MeV. Applying this constraint at the $\nu=1$ data we get $\lambda \ge 0.988$, which means that the majority of pressure at $T=155$~MeV comes from the fermions if we require physical $p/\epsilon$ values for the gluon gas at the same temperature. With the choice $\lambda=0.988$, the $p/\epsilon$ plots in Fig.~(\ref{fig:fig6}) are derived. We note that for all exponents $\nu$ studied, $p/\epsilon$ tends asymptotically to the relativistic limit for the full quasiparticle model as well as for the fermionic content of it. The fermion $p/\epsilon$ plots follow closely the full quasiparticle behavior and the initial condition seems to affect the curves only for $T$ near 155~MeV. On the contrary, the behavior of the gluon plots is strongly affected by the initial condition and only after ~250 MeV seems to stabilize.

\subsection{Two-flavor model}
\begin{figure}[htb]
\centerline{\includegraphics[width=\columnwidth]{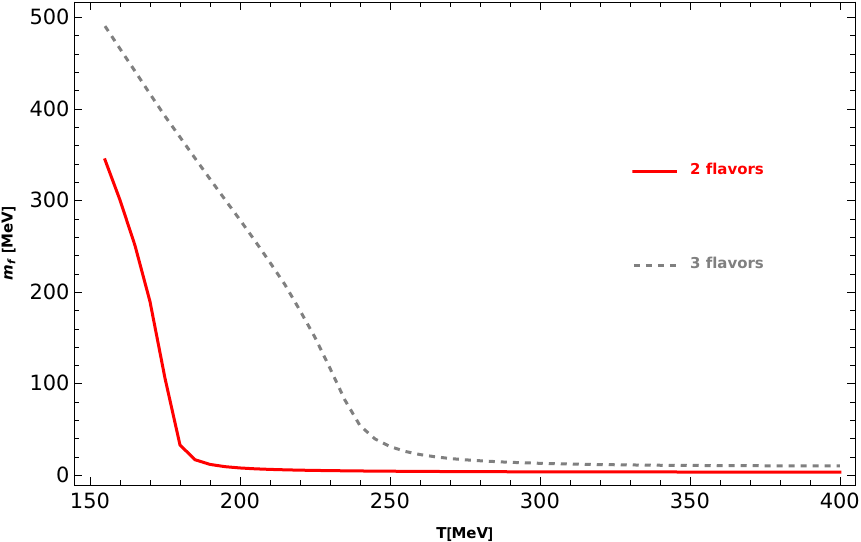}}
\vspace{3mm}
\centerline{\includegraphics[width=\columnwidth]{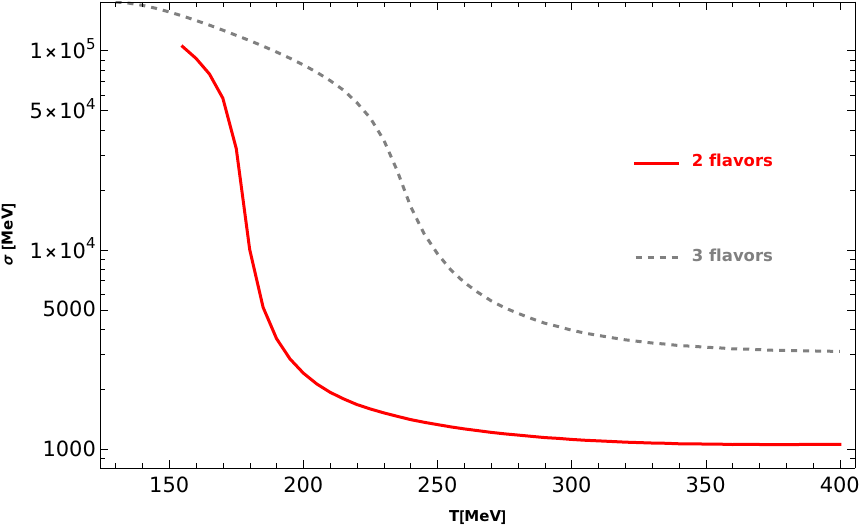}}
\caption{(a) The temperature dependence of the quark mass, $m_f(T)$ for the two-flavor ENAQPM in the $[155, 400]$ MeV interval (red line) versus the three-flavor system (grey line) for monoparametric fits of the energy density assuming a linear 
relation $\sigma(T) \sim m_f(T)$.  (b) The temperature-dependence of the gluon variance, $\sigma(T)$, for the two-flavor model (red dots) versus the three-flavor system (grey dots) from the linear relation in (a).
}
\label{fig:fig7}
\end{figure}

\noindent
Since the three-flavor symmetry is only approximate in nature, we examined also an ENAQP model (Eq.~(\ref{eq:eq3})) with two degenerate flavors of fermionic matter. Although the Lattice QCD data refer to a (2+1)-flavor system with two light and one heavy quark, we consider useful to examine the influence of the heavy flavor in the thermodynamical behavior as the effects of the strange quark are generically suppressed due to its mass. 
Overall, this model has no noticeable qualitative differences to the ENAQP model with three identical quark flavours presented in the previous subsections. The monoparametric fit of the energy density is achieved equally well with the fermion mass obtaining a constituent value near 340~MeV at $ T =155$ ~MeV and decreasing rapidly to a light mass of 10~MeV at the high temperature limit of 400~MeV (Fig.~\ref{fig:fig7}(a)). We stress that the 340~MeV value at $T=155$~MeV is a result of the fit for $m_f(T)$ with the fixed value of 10~MeV at the $T=400$~MeV point.  This value is in accordance to constituent quark mass evaluations for the light flavors based on light quark systems. 
We notice also, a similar qualitatively behavior to the three-flavor system for the gluon variance (Fig.~\ref{fig:fig7}(b)).
In addition, we observe that the crossover behavior is more rapid with respect to the three-flavor model with the transition completed around 180~MeV. This is an effect in accordance to predictions based on effective models for the two-flavor massless system since a genuine second order transition is expected for the $\text{SU(2)}_L\times \text{SU(2)}_R \sim \text{O(4)}$-symmetric system (\cite{Wilczek1992}) and the transition turns to a crossover after adding a third matter component. We conclude that adding quark flavors with temperature-dependent mass to the non-abelian quasi-particle gluon model produces a thermodynamical behavior which agrees phenomenologically to observations based on flavor symmetry and actual Lattice QCD calculations.

\section{Concluding remarks}

\noindent
In this work we added quasi-particle quark flavors with a temperature-dependent mass to the non-abelian quasi-particle model for SU(3) YM-theory introduced in \cite{Politis2016}, with the goal to describe the thermodynamical behavior of QCD at zero chemical potential as obtained via the (2+1)-flavor Lattice QCD calculations. Both bosonic and fermionic components are non-interacting with all interaction and thermal effects assumed to be captured by the dominant temperature-dependent parameters.
These are the fermion quasi-particle mass, $m_f(T)$, in the fermionic sector and the gluon variance, $\sigma(T)$, in the bosonic sector. We assumed quasi-Gaussian distributions peaked at zero mass for the transverse and longitudinal gluon modes and thus the dominant parameter $\sigma(T)$ is directly proportional to the actual distribution average mass and its variance.  With this Extended Non-Abelian Quasi-Particle Model (ENAQPM), we demonstrated that an accurate description of the energy density of the Lattice QCD data is viable for temperatures in the $[155, 400]$~MeV range with a correlated behavior of the fermion and gluon parameters dictated by the fit. This correlation appears quite robust since a variety of relations with a range of exponents in $\sigma(T) \sim m_f(T)^\nu$ fit the data equally well. The quasi-particle fermion mass behavior deduced from the fit is phenomenologically interesting and compatible with an analytic connection between a hadronic (confinement) phase thermodynamically accessed via the Hadron Resonance Gas near 155~MeV and a liberated gas with quark mass at its light current value becoming valid after a temperature around 230~MeV. 
 At the same time , the rapid decrease of the gluon variance, observed for $155 < T < 230$~MeV,  is consistent with a rapid crossover to a phase with reduced gluon mass scales. Since their mass is reduced, in this phase the gluons propagate to larger distances with the same energy and this is a characteristic in accordance with the Quark-Gluon Plasma phase, traditionally named the `deconfinement' phase . Still, the energy density remains below the ideal Boltzmann gas limit in this phase and thus interactions remain important even at high temperatures. Most importantly, our model provides evidence that assuming quasi-particle prevalence in both the fermion and gluon sectors, chiral symmetry restoration for the quark flavors is interlinked to the deconfinement mechanism which liberates the color degrees of freedom with a continuous way between 155 and 230~MeV.  Thus, our model provides a simple tool for understanding at a general level the Lattice-QCD thermodynamics at zero chemical potential and possibly indicates a direction for the investigation of  the quark- gluon scales connection from a first-principles approach. It is interesting also that switching from three to two quark flavors in the quasi-particle model follows the trend of known phenomenological investigations, i.e. lighter constituent quark masses at the HRG matching scale and a sharper transition to the chirally symmetric/deconfined phase.  
 Within the quasi-particle picture it would be interesting to improve the model by deriving the gluon mass distribution from first principles, based on the quantum YM field-theoretical partition function near its classical Miskowskian saddle points.
Furthermore,  the inclusion of  chemical potential in the ENAQPM would enable investigations of the QCD phase diagram and in this case the model could also be utilized in a predictive manner. However, such an effort is left for a future study.





\end{document}